\documentclass[a4paper,12pt]{article}

\usepackage[dvips]{graphicx}
\usepackage{amsfonts}
\usepackage{amsmath,amssymb}
\usepackage{dsfont}

\usepackage[usenames]{color}

\newcommand{\heff}{h_{\text{eff}}}
\newcommand{\hbareff}{\hbar_{\text{eff}}}

\newcommand{\Int}{\int\limits}
\newcommand{\ud}{\text{d}}

\newcommand{\Z}{\mathbb{Z}}

\newcommand{\ui}{\text{i}}
\newcommand{\ue}{\text{e}}
\newcommand{\psireg}{\psi_{\text{reg}}}
\newcommand{\psich}{\psi_{\text{ch}}}

\newcommand{\Ureg}{U_{\text{reg}}}
\newcommand{\Uregt}{\widetilde{U}_{\text{reg}}}

\newcommand{\Preg}{P_{\text{reg}}}
\newcommand{\Pch}{P_{\text{ch}}}

\newcommand{\UT}{U_{\text{T}}}
\newcommand{\UVreg}{U_{\widetilde{V}}}
\newcommand{\UTreg}{U_{\widetilde{T}}}

\newcommand{\Hreg}{H_{\text{reg}}}
\newcommand{\Ereg}{E_{\text{reg}}}
\newcommand{\vchm}{v_{\text{ch},m}}

\newcommand{\mmax}{N_{\text{reg}}}
\newcommand{\A}{A}

\newcommand{\opone}{\mathds{1}}
\newcommand{\phireg}{\varphi_{\text{reg}}}

\newcommand{\Qdyn}{Q_{\text{dyn}}}
\newcommand{\Qdir}{Q_{\text{dir}}}

\newcommand{\mapho}{\mathcal{D}_{\text{ho}}}
\newcommand{\mapdef}{\mathcal{D}_{\text{d}}}

\newcommand{\mapd}{\mathcal{D}}
\newcommand{\mapsi}{\mathcal{D}_{\text{rs}}}

\begin{document}

\title{Direct Regular-to-Chaotic Tunneling Rates Using the 
       Fictitious Integrable System Approach}

\author{Arnd B\"acker, Roland Ketzmerick, and Steffen L\"ock
\vspace*{0.2cm}
\\
{\scriptsize Institut f\"ur Theoretische Physik, Technische Universit\"at Dresden, 01062 Dresden, Germany}
}

\date{\today}

\maketitle

\begin{abstract}
\noindent
In systems with a mixed phase space, where regular and chaotic motion
coexists, regular states are coupled to the chaotic region
by dynamical tunneling.
We give an overview on the determination of
direct regular-to-chaotic tunneling rates 
using the fictitious integrable system approach.
This approach is applied to different kicked systems, including
the standard map,
and successfully compared with numerical data.

\noindent
\emph{This text corresponds to Chapter 6 of the book: Dynamical Tunneling - Theory and Experiment,
edited by S.~Keshavamurthy and P.~Schlagheck [Taylor and Francis CRC (2011)] \cite{BaeKetLoe2011}.
For a more extensive exposition see [Phys.~Rev.~E 82, 056208 (2010); arXiv:1009.0418v2].}
\end{abstract}

\section{Introduction}
\label{sec:introduction}

Tunneling is one of the most fundamental manifestations
of quantum mechanics.
For 1D systems the theoretical description is well established
by the Wentzel-Kramers-Brillouin (WKB) method and related approaches 
\cite{LanLif1979,GilPat1977}. However, for higher dimensional systems
no such simple description exists.
In these systems typically regular and chaotic motion
coexists and in the two-dimensional case
regular tori are absolute barriers to the motion.
Quantum mechanically, 
the eigenfunctions either concentrate
within the regular islands or in the chaotic sea,
as expected from the semiclassical eigenfunction hypothesis 
\cite{Per1973,Ber1977,Vor1979}. These eigenfunctions
are coupled by so-called dynamical tunneling \cite{DavHel1981} through 
the dynamically generated barriers in phase space.
In particular, this leads to a substantial enhancement of
tunneling rates between phase-space regions of regular motion due to the 
presence of chaotic motion, which was termed chaos-assisted tunneling
\cite{LinBal1990,BohTomUll1993,TomUll1994} .
Such dynamical tunneling processes are ubiquitous 
in molecular physics and were realized with microwave cavities 
\cite{DemGraHeiHofRehRic2000} and cold 
atoms in periodically modulated optical lattices
\cite{SteOskRai2001,HenHafBroHecHelMcKMilPhiRolRubUpc2001}.

In the quantum regime, $\heff \lesssim\A$, where the effective Planck constant 
$\heff$ is smaller but still comparable to the area of the regular island $\A$, 
the process of tunneling into the chaotic region 
is dominated by a \textit{direct regular-to-chaotic
tunneling} mechanism \cite{HanOttAnt1984,PodNar2003,SheFisGuaReb2006,BaeKetLoeSch2008}.
For the prediction of tunneling rates the fictitious integrable system approach
was introduced recently \cite{BaeKetLoeSch2008}.
It relies on a fictitious integrable system \cite{BohTomUll1993,PodNar2003} 
that resembles the regular 
dynamics within the island under consideration.
The approach has been applied to
quantum maps \cite{BaeKetLoeSch2008}, billiard systems 
\cite{BaeKetLoeRobVidHoeKuhSto2008}, and the annular microcavity 
\cite{BaeKetLoeWieHen2009}. 
In the semiclassical regime, $\heff \ll \A$, however,
the direct tunneling contribution alone is typically
not sufficient to describe the observed 
tunneling rates, because nonlinear resonance chains within
the regular island lead to resonance-assisted tunneling 
\cite{BroSchUll2001,BroSchUll2002,EltSch2005,SchEltUll2005}.
Recently, a combined prediction of dynamical tunneling rates from regular 
to chaotic phase-space regions was derived \cite{LoeBaeKetSch2010},
which combines the direct regular-to-chaotic tunneling mechanism
in the quantum regime with an improved resonance-assisted tunneling
theory in the semiclassical regime. 
In this text we concentrate on the fictitious integrable system approach
for the theoretical description of the direct regular-to-chaotic tunneling
mechanism and how it can be applied analytically, 
semiclassically, and numerically to quantum maps.

This text is organized as follows:
We start by defining different convenient kicked systems,
their quantization, and the numerical determination of tunneling
rates. In Sec.~\ref{sec:dirtun} we describe the approach to 
determine dynamical tunneling rates by using 
a fictitious integrable system. 
In Sec.~\ref{sec:ex} we apply this theory
to the previously introduced systems
and compare its results with numerical data.
We conclude with a brief summary.

\section{Kicked systems}
\label{sec:kicked}

Kicked systems are particularly suited to study classical and quantum 
effects appearing in a mixed phase space as they 
can be easily treated analytically and numerically. In contrast to 
time-independent systems, where at least two degrees of freedom are necessary to 
break integrability, one-dimensional driven systems can show chaotic motion. 
We consider time-periodic one-dimensional kicked systems 
\begin{eqnarray}
\label{eq:maps:kicked_hamiltonian}
 H(q,p,t) = T(p) + V(q) \sum_{n\in\Z} \tau \delta(t - n\tau),
\end{eqnarray}
which are described by the kinetic energy $T(p)$ and the potential $V(q)$
which is applied once per kick period $\tau=1$.
The classical dynamics of a kicked system is given by its stroboscopic mapping,
e.g., evaluated just after each kick
\begin{eqnarray}
\label{eq:maps:kicked_map}
 \begin{array}{l}
  q_{n+1}=q_{n}+T'(p_{n}),\\ 
  p_{n+1}=p_{n}-V'(q_{n+1}).
 \end{array}
\end{eqnarray}
It maps the phase-space coordinates after the $n$th kick to those
after the $(n+1)$th kick.
For the stroboscopic mapping we
consider a compact phase space with periodic 
boundary conditions for $q \in [-1/2, \, 1/2]$ and $p \in [-1/2, \, 1/2]$.
The phase space generally consists of regions 
with regular motion surrounded by chaotic 
dynamics, see Fig.~\ref{fig:maps:standard_map}(b). 
We focus on the situation of just one regular island embedded in
the chaotic sea, Fig.~\ref{fig:maps:designed}(a),(b). 
At the center of the island one has an elliptic fixed point, which is 
surrounded by invariant regular tori.
Classically, a particle within a regular island will never enter
into the chaotic region and vice versa, i.e.\ tori form absolute barriers to 
the motion.

\subsection{Standard map}
\label{sec:kicked:standard}

\begin{figure}[t]
  \begin{center}
    \includegraphics[]{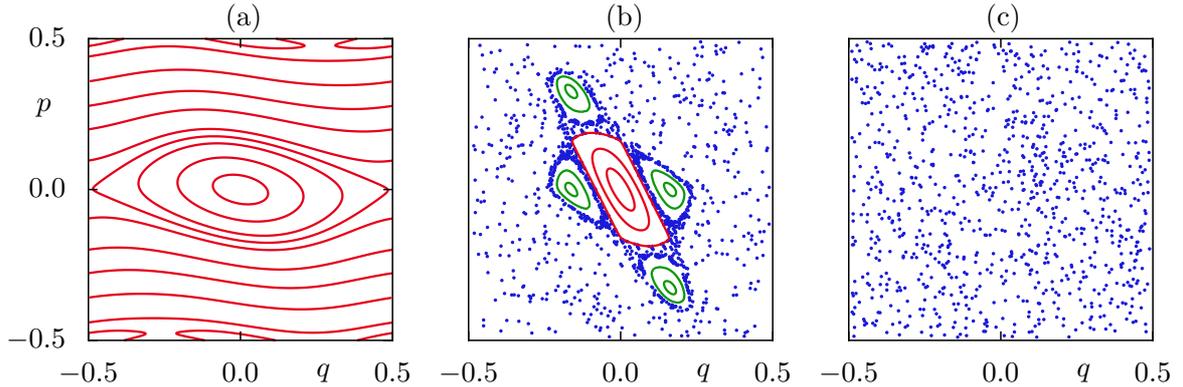}
    \caption{Phase space of the standard map for (a) $\kappa=0.3$, (b) 
             $\kappa=2.4$, and (c) $\kappa=8$. In (a) the system is nearly
             integrable. Most of the phase space is filled by invariant tori 
             (lines).
             The second figure (b) shows a mixed phase space with regular 
             motion (lines) and chaotic motion (dots). 
             The four small islands surrounding the central island form a 
             nonlinear resonance chain. In the chaotic sea partial barriers 
             exist, which is reflected by the different densities of 
             chaotic points. For large values of
             $\kappa$ as in (c) the phase space seems macroscopically chaotic.
           }
    \label{fig:maps:standard_map}
  \end{center}
\end{figure}

The paradigmatic model of an area preserving map is the standard map, defined 
by Eq.~\eqref{eq:maps:kicked_map} with the functions
\begin{eqnarray}
\label{sec:maps:stmap:TV}
 T'(p) & = & p, \\
 V'(q) & = & \frac{\kappa}{2\pi}\sin(2\pi q).
\end{eqnarray}
For $\kappa=0$ the system is integrable and the dynamics takes place on 
invariant tori with constant momentum. 
For small $\kappa > 0$, see Fig.~\ref{fig:maps:standard_map}(a), the system 
is no longer integrable. However, as stated by the Kolmogorov-Arnold-Moser (KAM) 
theorem \cite{Ber1978}, many invariant tori persist. In between these tori 
sequences of stable and unstable fixed points emerge, which is described 
by the Poincar\'e-Birkhoff theorem \cite{ArrPla1990}. The stable fixed points 
of such a sequence form so-called nonlinear resonance chains, 
see Fig.~\ref{fig:maps:standard_map}(b), while in the vicinity of 
the unstable fixed points typically chaotic dynamics is observed. 
For larger $\kappa$
more and more regular tori break up and the chaotic dynamics 
occupies a larger area of phase space. At $\kappa\approx 3$ 
one large regular region (``regular island'') is
surrounded by a region of chaotic dynamics (``chaotic sea'').
At the border of the regular island a hierarchical region exists,
where self-similar structures of regular and chaotic dynamics are 
found on all scales.
Also in the chaotic sea additional structures can be found. Here,
partial barriers limit the classical flux between different
regions of chaotic motion caused by the stable and unstable manifolds of 
unstable periodic orbits or caused by cantori, which are the remnants of 
regular tori \cite{KayMeiPer1984b}.
For even larger $\kappa$ the regular islands disappear and the motion
seems macroscopically chaotic, see Fig.~\ref{fig:maps:standard_map}(c). 

All these structures in phase space, such as nonlinear resonances, 
the hierarchical regular-to-chaotic
transition region, and partial barriers within the chaotic sea, 
have an influence on the dynamical 
tunneling process from the regular island to the chaotic sea. 
In order to quantitatively understand dynamical tunneling in systems with a 
mixed phase space, it is helpful to first consider simpler model systems. 
By designing maps such that the phase-space structures can be switched
on and off one-by-one it is possible to study their influence on the dynamical 
tunneling process separately. Afterwards the results can be tested using the 
generic example of the standard map.

\subsection{$\mapd$esigned maps}
\label{sec:kicked:designed}

Our aim is to introduce kicked systems which are designed such that
their phase space is particularly simple. It shows one regular island embedded 
in the chaotic sea, with 
very small nonlinear resonance chains within the regular island, a narrow 
hierarchical region, and without relevant partial barriers in the chaotic 
component. For such a system it is possible to study the direct 
regular-to-chaotic
tunneling process without additional effects caused by these structures
as long as $\heff$ is big enough.
To this end we define the family of maps $\mapd$, according to 
Eq.~\eqref{eq:maps:kicked_map},
with an appropriate choice of the functions $T'(p)$ and $V'(q)$ \cite{SchOttKetDit2001,BaeKetMon2005,BaeKetLoeSch2008,LoeBaeKetSch2010}.
For this we first introduce
\begin{eqnarray}
\label{sec:maps:amphib:tv_disc}
 t'(p) = \frac{1}{2} \pm (1-2p) &\;\text{for}\;& 0 < \pm p < \frac{1}{2}, \\
 v'(q) = -rq+Rq^2              &\;\text{for}\;&  -\frac{1}{2} < q < \frac{1}{2}
\end{eqnarray}
with $0<r<2$ and $R \geq 0$.
Considering periodic boundary conditions the functions
$t'(p)$ and $v'(q)$ show discontinuities at $p=0,\pm 1/2$ and
$q=\pm 1/2$, respectively. In order to avoid these discontinuities we smooth 
the periodically extended functions $v'(q)$ and $t'(q)$ with a Gaussian, 
\begin{equation}
\label{sec:maps:amphib:gaussian}
  G(z)=\frac{1}{\sqrt{2\pi\varepsilon^2}}\exp\left(-\frac{z^2}{2\varepsilon^2} 
                \right),
\end{equation}
resulting in analytic functions 
\begin{eqnarray}
\label{sec:maps:amphib:tv_cont}
 T'(p) & = & \Int \ud z \, t'(z)G(p-z),\\  
 V'(q) & = & \Int \ud z \, v'(z)G(q-z),
\end{eqnarray}
which are periodic with respect to the phase-space unit cell. With this we 
obtain the maps $\mapd$ depending on the 
parameters $r$, $R$, and the smoothing strength $\varepsilon$. 
The smoothing $\varepsilon$ determines the size of the hierarchical region
at the border of the regular island. Tuning the parameters $r$ and $R$ one 
can find situations, where all nonlinear resonance chains inside the regular 
island are small.

\begin{figure}[t]
  \begin{center}
    \includegraphics[]{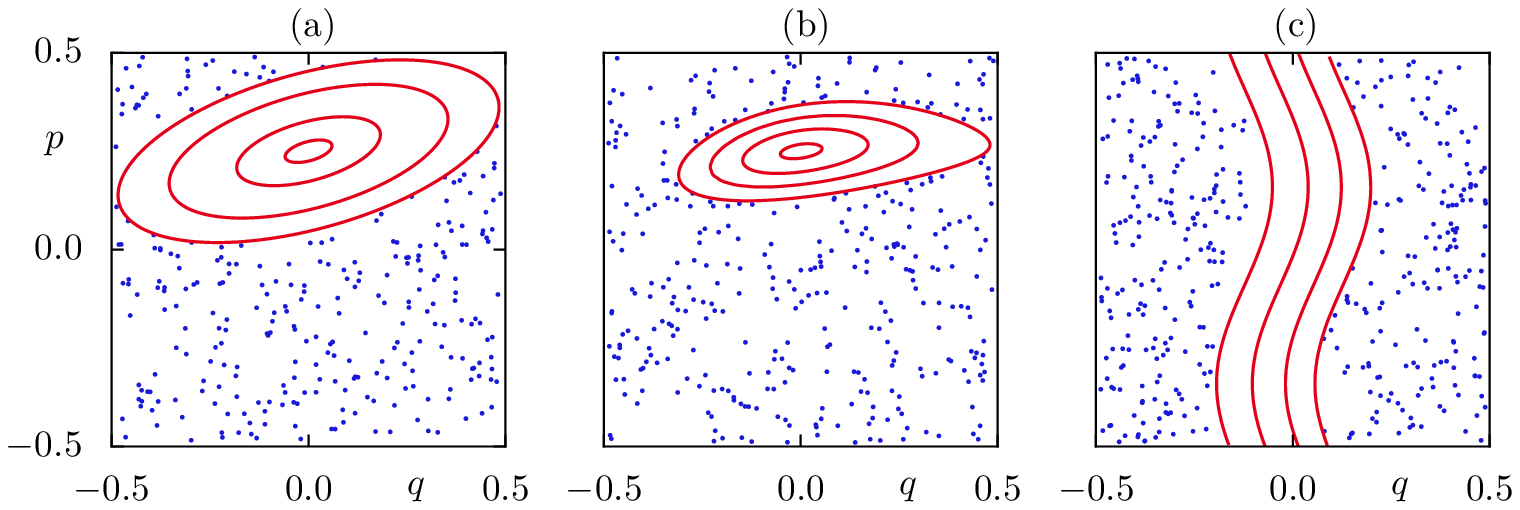}
    \caption{Phase space of (a) the designed map $\mapho$ with a harmonic 
             oscillator-like regular island, (b) $\mapdef$ with a deformed 
             regular island, and (c) $\mapsi$ with a regular stripe. For 
             parameters see text.
           }
    \label{fig:maps:designed}
  \end{center}
\end{figure}

For $R=0$ both functions $v'(q)$ and $t'(p)$ are linear in $q$ and $p$,
respectively. In this case we find a harmonic oscillator-like regular 
island with elliptic invariant tori and constant rotation number. 
We choose the parameters $r=0.46$, $R=0$, and $\varepsilon=0.005$
for which the phase space of the resulting map $\mapho$
is shown in Fig.~\ref{fig:maps:designed}(a).

In typical systems the rotation number of regular tori changes from the center 
of the regular region to its border which typically has a non-elliptic shape. 
Such a situation
can be achieved for the family of maps $\mapd$ with the parameter $R\neq 0$.
For most combinations of the parameters $r$ and $R$ resonance structures
appear inside the regular island. They limit the $\heff$-regime in which the
direct regular-to-chaotic tunneling process dominates. Hence, we choose a 
situation in which the nonlinear resonances are small such that
their influence on the tunneling process is expected only at small $\heff$.
The phase space of the map $\mapdef$, obtained
for $r=0.26$, $R=0.4$, and $\varepsilon=0.005$,
is shown in Fig.~\ref{fig:maps:designed}(b).

Another designed kicked system was introduced in Refs.~\cite{ShuIke1995,ShuIke1998,IshTanShu2007}. 
Here the regular region consists of a stripe in phase space, see 
Fig.~\ref{fig:maps:designed}(c). 
In our notation the mapping, Eq.~\eqref{eq:maps:kicked_map}, is specified by 
the functions 
\begin{eqnarray}
 \label{eq:maps:shudo:v}
 V'(q) & = & -\frac{1}{2\pi}\left(8\pi aq+d_1-d_2+\frac{1}{2}
             [8\pi aq-\omega+d_1] \tanh[b(8\pi q-q_d)]\right.\nonumber\\
    &   & \left. +\frac{1}{2}[-8\pi aq+\omega+d_2]\tanh[b(8\pi q+q_d)]\right)\\
 \label{eq:maps:shudo:t}
 T'(p) & = & -\frac{\kappa}{8\pi}\sin(2\pi p).
\end{eqnarray}
The kinetic energy $T(p)$ is periodic with respect to the phase space unit 
cell. We label the resulting map with a regular stripe by $\mapsi$
using the parameters $a=5$, $b=100$, 
$d_1=-24$, $d_2=-26$, $\omega = 1$, $q_d = 5$, and $\kappa = 3$.
 
The map $\mapsi$ is similar to the system $\mapho$
as it also destroys the integrable region by smoothly 
changing the function $V'(q)$, here at $|q|=q_d/(8\pi)$. 
For $|q| < q_d/(8\pi)$ the potential term is almost linear while it tends to 
the standard map for $|q| > q_d/(8\pi)$. The parameter $b$ determines the width
of the transition region.

\subsection{Quantization}
\label{sec:kicked:quant}

Quantum mechanically kicked systems are described by the unitary 
time-evolution operator $U$ \cite{BerBalTabVor1979}. It can be written as 
$U=U_V U_T$ with
\begin{eqnarray}  
\label{eq:maps:quantum_map}
 U_V & = & \ue^{-\ui V(q)/\hbareff}, \\
 U_T & = & \ue^{-\ui T(p)/\hbareff}.
\end{eqnarray}
Its eigenstates $|\psi_n\rangle$ and quasi-energies $\varphi_n$ 
are determined by
\begin{eqnarray}
\label{eq:deriv:eigen_eq}
 U|\psi_n\rangle = \ue^{\ui\varphi_n}|\psi_n\rangle.
\end{eqnarray}
We consider a compact phase space with periodic boundary conditions in $q$ and
$p$. This implies that the effective Planck constant can take 
only the discrete values \cite{ChaShi1986}
\begin{eqnarray}
 \heff = \frac{1}{N},
\end{eqnarray}
where $N$ is the dimension of the Hilbert space. In addition the position and 
momentum coordinates are quantized according to
\begin{eqnarray}
 q_k & = & \frac{k}{N}-\frac{1}{2},\\
 p_l & = & \frac{l}{N}-\frac{1}{2}
\end{eqnarray}
with $k,l=0,\dots,N-1$.

\begin{figure}[b]
  \begin{center}
    \includegraphics[]{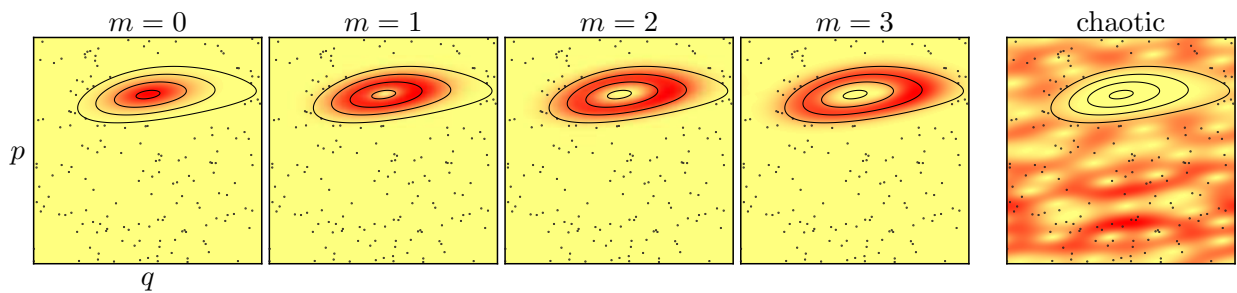}
    \caption{Husimi representation of eigenfunctions of the map $\mapdef$ 
             for $\heff=1/50$.
             The regular ground state $m=0$ and the first three excited states
             $m=1,2,3$ as well as a chaotic state are shown.
           }
    \label{fig:maps:eigenstates}
  \end{center}
\end{figure}

For systems with a mixed phase space, in the semiclassical
limit ($\heff \to 0$), the semiclassical eigenfunction hypothesis 
\cite{Per1973,Ber1977,Vor1979} implies that the eigenstates $|\psi_n\rangle$ 
can be classified as either regular or chaotic according to the phase-space
region on which they concentrate, see Fig.~\ref{fig:maps:eigenstates}.
In order to understand the behavior of eigenstates 
away from the semiclassical limit, i.e.\ at finite values of
the effective Planck constant $\heff$, one has to compare the size of
phase-space structures with $\heff$. 

The so-called regular states are concentrated on tori within the regular island 
and fulfill the Bohr-Sommerfeld type quantization condition
\begin{eqnarray}
\label{eq:deriv:EBK}
 \oint p\,\ud q = \heff\left(m+\frac{1}{2}\right),\quad m=0,\dots,\mmax-1.
\end{eqnarray}
For a given value of $\heff$ there exist $\mmax$ of such 
regular states, where $\mmax=\lfloor \A/\heff+1/2 \rfloor$ and $A$ is the
area of the regular island. The $N_{\text{ch}}=N-\mmax$ chaotic states 
extend over the chaotic sea. Due to dynamical tunneling, however,
the regular and chaotic eigenfunctions of $U$ always 
have a small component in the other region of phase space, respectively. 
This is most clearly seen for hybrid states which even have the same weight 
in each of the components as they are involved in an avoided 
level crossing.

\subsection{Numerical determination of tunneling rates}
\label{sec:kicked:numerical}

The structure of the considered phase space, with one regular island 
surrounded by the chaotic sea, allows for the determination of tunneling rates 
by introducing absorption somewhere in the chaotic region of phase space. 
For quantum maps, this can for example be realized by using a non-unitary
open quantum map \cite{SchTwo2004}
\begin{eqnarray}
\label{eq:numerics:open_Uo}
 U^o = PUP,
\end{eqnarray}
where $P$ is a projection operator onto the complement of the absorbing region. 

While the eigenvalues of $U$ are located on the unit circle the eigenvalues 
of $U^o$ are inside the unit circle as $U^o$ is sub-unitary. 
The eigenequation of 
$U^o$ reads
\begin{eqnarray}
 U^o|\psi_{n}^{s}\rangle = z_n|\psi_{n}^{s}\rangle
\end{eqnarray}
with eigenvalues
\begin{eqnarray}
\label{eq:numerics:open_evals}
 z_n = \ue^{\ui\left(\varphi_n + \ui\frac{\gamma_n}{2} \right)}.
\end{eqnarray}
The decay rates are characterized by the imaginary part of the quasi-energies 
in Eq.~\eqref{eq:numerics:open_evals} and one has
\begin{eqnarray}
\label{eq:numerics:open}
 \gamma_m = -2\log|z_m| \approx 2(1-|z_m|).
\end{eqnarray}

If the chaotic region does not contain partial barriers and shows no dynamical 
localization, it is justified to assume that the 
rate of escaping the regular island is equal to the rate of 
leaving through the absorbing regions located in the chaotic sea. 
Then, the location of the absorbing regions in the chaotic part of phase space
has no effect on the tunneling rates.

In generic systems, however, partial barriers will appear in the chaotic 
region of phase space. The additional transition through these structures 
further limits the quantum transport such that the calculated decay through 
the absorbing region occurs slower than the decay from the regular island to 
the neighboring chaotic sea. Similarly dynamical localization in the chaotic
region may slow down the decay. The influence of partial barriers and 
dynamical localization on the tunneling process is an open problem. Here we 
will suppress their influence, if necessary, by 
moving the absorbing regions closer to the regular island.

\begin{figure}[t]
 \begin{center}
  \includegraphics[]{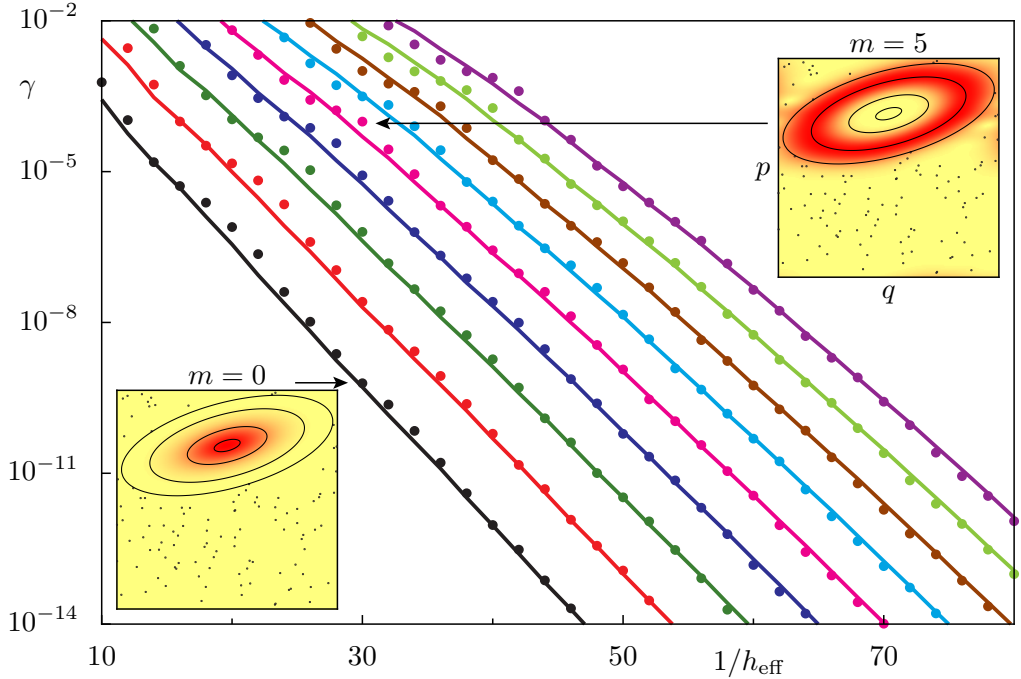}
  \caption[]
        {Numerical data (dots) for $m \leq 8$ 
         for the map $\mapho$ with a harmonic oscillator-like island.
         Comparison with Eq.~\eqref{eq:deriv:final_result} (lines),
         see Sec.~\ref{sec:dirtun:deriv} for its derivation. 
         The insets show Husimi functions of the regular states for the 
         quantum numbers $m=0$ and $m=5$ at $1/\heff=30$.}
         \label{fig:maps:rates_amph_ellipt}
 \end{center}
\end{figure}

In Fig.~\ref{fig:maps:rates_amph_ellipt} tunneling rates are determined 
numerically by opening the system for the map $\mapho$ (dots). These 
agree very well with an analytical prediction (lines), 
Eq.~\eqref{eq:deriv:final_result}, 
which is derived in the next section.

\section{Tunneling using a fictitious integrable system}
\label{sec:dirtun}

Dynamical tunneling in systems with a
mixed phase space couples the regular island and the 
chaotic sea, which are classically separated. 
This coupling can be 
quantified by tunneling rates $\gamma_m$ which describe the decay of regular 
states to the chaotic sea. To define these tunneling rates one can consider 
a wave packet started on the $m$th quantized torus in the regular island
coupled to a continuum of chaotic states, as in the case for an infinite 
chaotic sea or in the presence of an absorbing region somewhere in the chaotic 
sea. Its decay $\ue^{-\gamma_m t}$ is described by a tunneling rate $\gamma_m$.
For systems with a finite phase space this exponential decay occurs at most up 
to the Heisenberg time $\tau_H=\heff/\Delta_{\text{ch}}$, where 
$\Delta_{\text{ch}}$ is the mean level spacing of the chaotic states. 

In the quantum regime,
$\heff \lesssim A$, where $\heff$ is smaller but comparable to the area $A$ of 
the regular island, the rates 
$\gamma_m$ are dominated by the direct regular-to-chaotic tunneling mechanism, 
while contributions from resonance-assisted tunneling are negligible.
We concentrate on situations where additional phase-space structures within 
the chaotic sea are not relevant for tunneling.
In the following we derive a prediction for the direct regular-to-chaotic 
tunneling rates using the fictitious integrable system approach \cite{BaeKetLoeSch2008}.

\subsection{Derivation}
\label{sec:dirtun:deriv}

\begin{figure}[t]
  \begin{center}
    \includegraphics[]{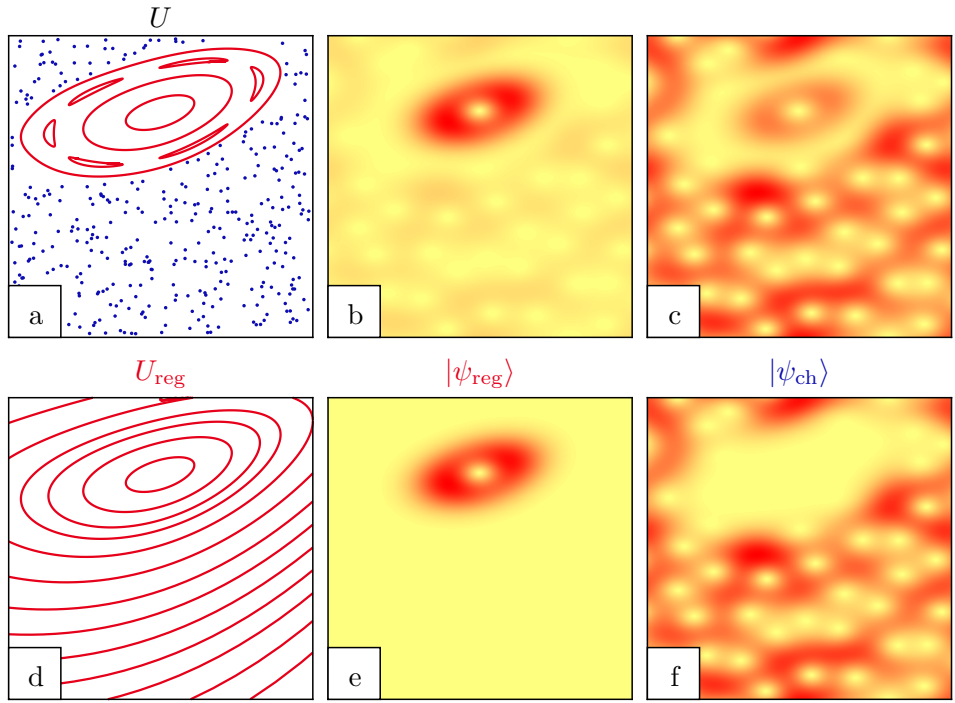}
    \caption{(a) Illustration of the classical phase 
          space corresponding to some mixed quantum map $U$ 
          together with the Husimi representation of (b) a regular and (c)
          a chaotic eigenstate of $U$ which both have a small component in the
          other region. (d) Illustration of the classical phase space of the 
          fictitious integrable system $\Ureg$. (e) Eigenstates 
          $|\psireg\rangle$ of $\Ureg$ are purely regular while (f) the basis 
          states $|\psich\rangle$ are localized in the chaotic region of 
          phase space.
          }
    \label{fig:deriv:quantum_maps}
  \end{center}
\end{figure}

In order to find a prediction for the direct regular-to-chaotic tunneling rates
we decompose the Hilbert space of the quantum map $U$ into two parts which
correspond to the regular and chaotic regions.
While classically such a decomposition is unique (neglecting tiny phase-space
structures), quantum mechanically this is not the case due to the uncertainty 
principle. We find a decomposition by introducing a fictitious integrable 
system $\Ureg$ (a related idea was presented in 
Refs.~\cite{BohTomUll1993,PodNar2003,SheFisGuaReb2006}). 
It has to be chosen such that its dynamics resembles 
the classical motion corresponding to $U$ within the regular island as closely 
as possible and continues this regular dynamics beyond the regular island of 
$U$, see Fig.~\ref{fig:deriv:quantum_maps}(a),(d). 
The eigenstates $|\psireg^m\rangle$ of $\Ureg$ are purely regular in 
the sense that they are localized on the $m$th quantized torus 
of the regular region and continue to decay beyond this regular region, 
see Fig.~\ref{fig:deriv:quantum_maps}(e). This is the decisive property of
$|\psireg^m\rangle$ which have no chaotic admixture, 
in contrast to the predominantly regular eigenstates of $U$, 
see Fig.~\ref{fig:deriv:quantum_maps}(b). 
The explicit construction of $\Ureg$ is 
discussed in Sec.~\ref{sec:dirtun:regsyst}.

With the eigenstates $|\psireg^m\rangle$ of $\Ureg$, 
$\Ureg|\psireg^m\rangle = \ue^{\ui \phireg^m}|\psireg^m\rangle$, we define a 
projection operator
\begin{equation}
\label{eq:deriv:P}
 \Preg := \sum_{m=0}^{\mmax-1} |\psireg^m\rangle\langle\psireg^m|,
\end{equation}
using the first $\mmax$ regular states of $\Ureg$ which approximately projects onto 
the regular island corresponding to $U$.
The orthogonal projector 
\begin{equation}
\label{eq:deriv:Pch}
 \Pch := \opone-\Preg 
\end{equation}
approximately projects onto the chaotic phase-space region. These projectors, 
$\Preg$ and $\Pch$, define our decomposition of the Hilbert space into 
a regular and a chaotic subspace.
 
Introducing an orthonormal basis $|\psich\rangle$ in the chaotic subspace we can write
$\Pch = \sum_{\text{ch}} |\psich\rangle\langle\psich|$. Here we sum over all 
$N_{\text{ch}}=N-\mmax$ states $|\psich\rangle$, 
see Fig.~\ref{fig:deriv:quantum_maps}(f) for an example.
Hence, the coupling matrix element $\vchm$ between a purely regular state 
$|\psireg^m\rangle$ and any chaotic basis state $|\psich\rangle$ is
\begin{equation}
\label{eq:deriv:coupmatelorth}
 \vchm = \langle\psich|U|\psireg^m\rangle.
\end{equation}
The tunneling rate is obtained using a dimensionless version of Fermi's 
golden rule,
\begin{eqnarray}
\label{eq:deriv:FGR}
 \gamma_m = \sum_{\text{ch}}|\vchm|^2,
\end{eqnarray}
where the sum is over all chaotic basis states $|\psich\rangle$
and thus averages the modulus squared of the fluctuating matrix elements $\vchm$. 
Here we apply Fermi's golden rule in the case of a discrete spectrum, 
which is possible if one considers the decay $\ue^{-\gamma_m t}$
up to the Heisenberg time $\tau_H=\heff/\Delta_{\text{ch}}$ only.
Inserting Eq.~\eqref{eq:deriv:coupmatelorth} we obtain
\begin{eqnarray}
\label{eq:deriv:final_result}
 \gamma_m = \Vert\Pch U|\psireg^m\rangle\Vert^2 = 
            \Vert (\opone-\Preg)U|\psireg^m\rangle\Vert^2
\end{eqnarray}
as the basis of all our following investigations. 
It allows for the prediction of tunneling rates 
from a regular state localized on the $m$th quantized torus to the chaotic sea.
Equation~\eqref{eq:deriv:final_result} confirms the intuition that the 
tunneling rates are determined by the amount of probability of 
$|\psireg^m\rangle$ that is transferred to the chaotic region after 
one application of the time evolution operator $U$.
In Eq.~\eqref{eq:deriv:final_result} properties of the fictitious 
integrable system $\Ureg$ and the chaotic projector $\Pch=\opone-\Preg$ enter,
which rely on the chosen decomposition of Hilbert space. 

In cases where one finds a fictitious integrable system $\Ureg$ which
resembles the dynamics within the regular island of $U$ with very high 
accuracy, Eq.~\eqref{eq:deriv:final_result} can be approximated as
\begin{eqnarray}
\label{eq:deriv:final_result_app}
 \gamma_m & \approx & \Vert (U-\Ureg)|\psireg^m\rangle \Vert^2,
\end{eqnarray}
using $\Preg U|\psireg^m\rangle \approx \Preg\Ureg|\psireg^m\rangle= 
\Ureg|\psireg^m\rangle$.
Instead of the operator product $\Pch U$ in Eq.~\eqref{eq:deriv:final_result}
the difference $U-\Ureg$ enters in Eq.~\eqref{eq:deriv:final_result_app}.
It allows for further derivations, which are presented in 
Sec.~\ref{sec:dirtun:approx}.

\subsection{Fictitious integrable system}
\label{sec:dirtun:regsyst}

The most difficult step in the application of  
Eq.~\eqref{eq:deriv:final_result} to a given system
is the determination of the fictitious integrable system $\Ureg$. 
On the one hand its dynamics should resemble the classical motion 
of the considered mixed system within the regular island as closely 
as possible.
As a result the contour lines of the corresponding integrable Hamiltonian 
$\Hreg$, Fig.~\ref{fig:deriv:quantum_maps}(d), 
approximate the KAM-curves of the 
classically mixed system, Fig.~\ref{fig:deriv:quantum_maps}(a), in phase space.
This resemblance is not possible with arbitrary precision as the integrable 
approximation does not contain, e.g., nonlinear resonance chains and small 
embedded chaotic regions. Moreover, it cannot account for the hierarchical 
regular-to-chaotic transition region at the border of the regular island.
Similar problems appear for the analytic continuation of a regular torus into 
complex space due to the existence of natural boundaries 
\cite{Wil1986, Cre1998, ShuIke1995, ShuIke1998, OniShuIkeTak2001, BroSchUll2001, 
BroSchUll2002, EltSch2005}.
However, for not too small $\heff$, where 
these small structures are not yet resolved, an integrable approximation with 
finite accuracy turns out to be sufficient for a prediction of the tunneling 
rates.

On the other hand the integrable dynamics of $\Hreg$ should extrapolate 
smoothly beyond the regular island of $H$. This is essential for the quantum 
eigenstates of $\Hreg$ to have correctly decaying
tunneling tails which are according to Eq.~\eqref{eq:deriv:final_result} 
relevant for the determination of the tunneling rates. 
While typically tunneling from the regular island occurs to 
regions within the chaotic sea close to the border of the regular island, there 
exist other cases, where it occurs to regions deeper inside the chaotic sea, as
studied in Ref.~\cite{SheFisGuaReb2006}. Here $\Hreg$ has to be constructed 
such that its eigenstates have the correct tunneling tails up to this region.

For quantum maps we determine the fictitious integrable system in the following 
way: We employ classical methods, see below, to obtain a one-dimensional 
time-independent Hamiltonian $\Hreg(q,p)$
which is integrable by definition and resembles the classically regular motion
corresponding to the mixed system.
After its quantization we obtain the regular quantum map 
$\Ureg=\ue^{-\ui\Hreg/\hbareff}$ which has the same
eigenfunctions $|\psireg^m\rangle$ as $\Hreg$. For the numerical evaluation of  
Eq.~\eqref{eq:deriv:final_result} we use 
$\Pch = \opone - P_{\text{reg}}=\opone-\sum|\psireg^m\rangle\langle\psireg^m|$,
according to Eqs.~\eqref{eq:deriv:P} and \eqref{eq:deriv:Pch},
where the sum extends over $m=0,\,1,\,\dots,\,\mmax-1$.

Two examples for the explicit construction of $\Hreg$ will be mentioned below.
Note, that also other methods, e.g.\ based on the normal-form analysis 
\cite{Gus1966,BazGioSerTodTur1993} 
or on the Campbell-Baker-Hausdorff formula \cite{Sch1988} 
can be employed in order to find $\Hreg$. For the example systems considered 
in this manuscript, however, they show less good agreement.

\begin{figure}[t]
  \begin{center}
    \includegraphics[]{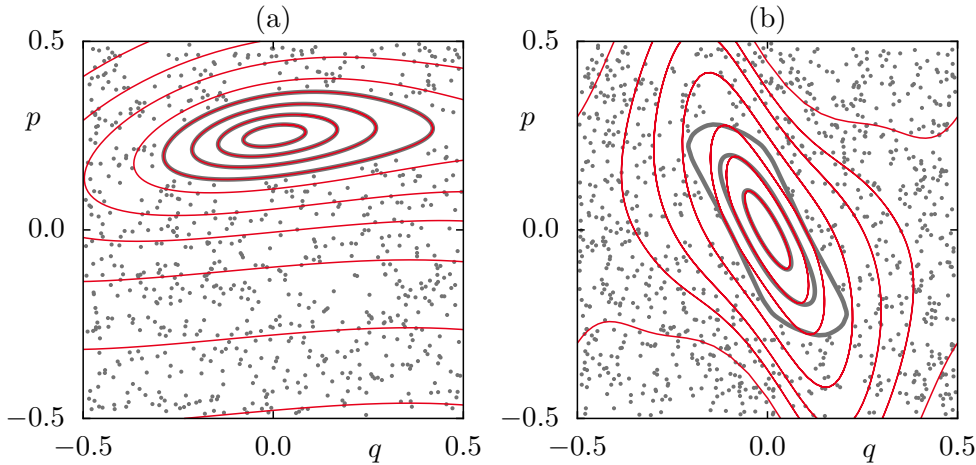}
    \caption{Application of the Lie transformation method: 
            (a) Orbits (gray) of the map $\mapdef$  
            and of the corresponding 
            integrable system (thin lines) of order $K=15$. Here $\Hreg$ accurately 
            resembles the regular dynamics of $U$. 
            (b) Orbits of the standard map for $\kappa=2.9$ (gray) and of the 
            corresponding integrable system (thin lines) of order $K=7$.
            Here $\Hreg$ does not accurately resemble the regular dynamics of 
            $H$.}
    \label{fig:maps:comp_map_Hreg_lie}
  \end{center}
\end{figure}

One possible choice for the determination of the fictitious integrable system
for quantum maps is the Lie-transformation method 
\cite{LicLie1983}. It determines a classical Hamilton function as a series,
\begin{equation}
 \label{eq:deriv:hamiltonian_lie}
  \Hreg^K(q,p) = \sum_{l=1}^{K} h_{l}(q,p),
\end{equation}
see Ref.~\cite{BroSchUll2002} and Fig.~\ref{fig:maps:comp_map_Hreg_lie}(a) for 
an example.
Typically, the order of the expansion $K$ can be increased up to $20$ within 
reasonable numerical effort. The Lie-transformation method provides a
regular approximation $\Hreg$ which interpolates the dynamics inside the 
regular region and gives a smooth continuation into the chaotic sea. At some 
order $K$ the series should diverge due to the nonlinear resonances
inside the regular island. 
For strongly perturbed systems, such as the standard map at $\kappa>2.5$,
the Lie-transformation method may not be able to reproduce the regular dynamics 
of $U$, see Fig.~\ref{fig:maps:comp_map_Hreg_lie}(b).
Here we use a method \cite{BaeKetLoeSch2008} based on the frequency 
map analysis \cite{LasFroCel1992}.

An important question is whether the direct tunneling rates obtained
using Eq.~\eqref{eq:deriv:final_result}
depend on the actual choice of $\Hreg$ and how these results 
converge depending on the order $K$ of its perturbation series.
Ideally one would like to use classical measures, which describe the deviations 
of the regular system $\Hreg$ from the originally mixed one, to predict the 
error of Eq.~\eqref{eq:deriv:final_result} for the tunneling rates. 
However, these classical measures can only account 
for the deviations within the regular region but not for the quality of the 
continuation of $\Hreg$ beyond the regular island of $U$. 
Currently, the quality of an  
integrable system can be estimated a posteriori by comparison of the predicted 
tunneling rates with numerical data.
It remains an open question how to obtain a direct connection between the 
error on the classical side and the one for the tunneling rates.

\subsection{Approximate fictitious integrable system}
\label{sec:dirtun:approx}

For an analytical evaluation of the result, 
Eq.~\eqref{eq:deriv:final_result}, of the direct regular-to-chaotic 
tunneling rates we approximate the fictitious integrable system $\Ureg$ by
a kicked system, $\Uregt = \UVreg U_T$ or $\Uregt = U_V\UTreg$ with
\begin{eqnarray}  
\label{eq:maps:quantum_map_reg}
 \UVreg & = & \ue^{-\ui \widetilde{V}(q)/\hbareff}, \\
 \UTreg & = & \ue^{-\ui \widetilde{T}(p)/\hbareff}.
\end{eqnarray}
Here the functions $\widetilde{V}(q)$ and $\widetilde{T}(p)$ are a low
order Taylor expansion of $V(q)$ and $T(p)$, respectively, around the center 
of the regular island.
Note, that the classical dynamics corresponding to $\Uregt$ is typically not 
completely regular. Still the following derivation is applicable if
$\Uregt$ has the following properties:
(i) Within the regular island it has an almost identical classical dynamics 
as $U$, including nonlinear resonances and small embedded chaotic regions.
(ii) It shows mainly regular dynamics for a sufficiently wide region 
beyond the border of the regular island of $U$. 

Now we consider the specific case $\Uregt = \UVreg U_T$ 
and assume that both properties 
(i) and (ii) are fulfilled. As the dynamics of $U$ and $\Uregt$ are almost 
identical within the regular island of $U$, the approximate result, 
Eq.~\eqref{eq:deriv:final_result_app}, can be applied with $\Ureg$
replaced by $\Uregt$, giving
\begin{eqnarray}
\label{eq:deriv:final_result_app_2}
 \gamma_m & \approx & \Vert (U - \UVreg U_T)|\psireg^m\rangle\Vert^2 \\
          & = & \Vert (U U_{T}^{\dagger}\UVreg^{\dagger}-\opone) 
                \UVreg U_T |\psireg^m\rangle\Vert^2.
\end{eqnarray}
We now use that $|\psireg^m\rangle$, which is an eigenstate of the exact 
$\Ureg$ and $\Hreg$, is an approximate eigenstate of $\UVreg U_T$, leading to 
$\UVreg U_T|\psireg^m\rangle \approx 
\ue^{\ui \varphi_{\text{reg}}^{m}}|\psireg^m\rangle$.
With this we obtain 
\begin{eqnarray}
\label{eq:maps:sc:tunneling_rate_1b}
\gamma_m & \approx & \Vert (U_V \UVreg^{\dagger}-\opone) | 
                     \psireg^m \rangle\Vert^2. 
\end{eqnarray}
In position representation this reads
\begin{eqnarray}
  \label{eq:maps:sc:tunneling_rate_2}
    \gamma_m & \approx &
      2 \sum_{k=0}^{N-1} \left\vert \psireg^m (q_k)  \right
      \vert^{2} \left[ 1-\cos\left( \frac{\Delta V(q_k)}{\hbareff}
      \right)  \right],
\end{eqnarray}
where $\Delta V(q):=V(q)-\widetilde{V}(q)$ and $q_k = k/N-1/2$.
In the semiclassical limit the sum in Eq.~\eqref{eq:maps:sc:tunneling_rate_2} 
can be replaced by an integral
\begin{eqnarray}
  \label{eq:maps:sc:tunneling_rate_2b}
    \gamma_m & \approx & 
      2 \Int_{-1/2}^{1/2} \ud q \left\vert \psireg^m (q)  \right
      \vert^{2} \left[ 1-\cos\left( \frac{\Delta V(q)}{\hbareff}
      \right)  \right],
\end{eqnarray}
where we integrate over the whole position space $q \in [-1/2, \, 1/2]$.
Note, that for the complementary situation, where $\Ureg\approx U_V\UTreg$ is 
used in Eq.~\eqref{eq:deriv:final_result_app}, a similar result can be obtained 
in momentum representation
\begin{eqnarray}
  \label{eq:maps:sc:tunneling_rate_2c}
    \gamma_m & \approx & 
      2 \Int_{-1/2}^{1/2} \ud p \left\vert \psireg^m (p)  \right
      \vert^{2} \left[ 1-\cos\left( \frac{\Delta T(p)}{\hbareff}
      \right)  \right],
\end{eqnarray}
with $\Delta T(p):=T(p)-\widetilde{T}(p)$.

We now use a WKB expression for the regular states $|\psireg^m\rangle$. 
For simplicity we restrict to the case $\Hreg=p^2/2+W(q)$ leading to
\begin{eqnarray}
 \label{eq:maps:sc:wkb_wf_1d_ort}
  \psireg^m(q) & \approx & \sqrt{\frac{\omega}{2\pi|p(q)|}} 
                  \exp\left(-\frac{1}{\hbareff}\Int_{q_{m}^{r}}^{q} 
                  |p(q')|\text{d}q' \right),
\end{eqnarray}
which is valid for $q>q_{m}^{r}$.
Here $q_{m}^{r}$ is the right classical turning point of the $m$th quantizing 
torus, $\omega$ is the oscillation frequency, and $p(q)=\sqrt{2(\Ereg^m-W(q))}$.
The eigenstates $\psireg^m(q)$ decay exponentially beyond the classical turning 
point $q_{m}^{r}$. The difference of the potential energies 
$\Delta V(q) = V(q)-\widetilde{V}(q)$ 
approximately vanishes within the regular region and increases beyond its
border to the chaotic sea. Hence, the most 
important contribution in Eq.~\eqref{eq:maps:sc:tunneling_rate_2b} arises 
near the left or the right border, $q_{b}^{l}$ or $q_{b}^{r}$, of the 
regular island. For $q>q_{b}^{r}$ we rewrite the regular states 
\begin{eqnarray}
\psireg^m(q) & \approx & \psireg^m\left(q_{b}^{r}\right) 
      \exp\left(-\frac{1}{\hbareff}\Int_{q_{b}^{r}}^{q} |p(q')|\,\ud q'\right)
      \sqrt{\frac{p(q_{b}^{r})}{p(q)}}\\
\label{eq:maps:sc:WKB_eval_3}
     & \approx & \psireg^m\left(q_{b}^{r}\right) \exp\left(-\frac{1}{\hbareff}
         (q-q_{b}^{r})|p(q_{b}^{r})|\right),
\end{eqnarray}
where in the last step we use $p(q)\approx p(q_{b}^{r})$ in the vicinity of the
border.

In order to use Eq.~\eqref{eq:maps:sc:tunneling_rate_2b} we split the 
integration interval in two parts, such that 
$\gamma_m = \gamma_{m}^{l}+\gamma_{m}^{r}$,
corresponding to the contributions from the left and the right.
We now approximate
\begin{eqnarray}
 \Delta V(q) \approx \left\{
 \begin{array}{ll}
   0 & ,\; q_{m}^{r} \leq q \leq q_{b}^{r} \\
   c_b (q-q_{b}^{r}) & ,\; q > q_{b}^{r}
 \end{array}
 \right.
\end{eqnarray}
with some constant $c_b$ and find
\begin{eqnarray}
\label{eq:maps:sc:tunnelrate_randformel_intnorm_a}
\gamma_{m}^{r} & \approx & 2\hbareff \vert\psireg^m(q_{b}^{r})\vert^2 
     \Int_{0}^{x_{\text{max}}} \ue^{-2x|p(q_{b}^{r})|} [1-\cos(c_b x)]\,\ud x\\
\label{eq:maps:sc:tunnelrate_randformel_intnorm}
       & \approx & \frac{I\heff}{\pi} \vert\psireg^m(q_{b}^{r})\vert^2, 
\end{eqnarray}
where $x = (q-q_{b}^{r})/\hbareff$ and 
$x_{\text{max}} = (1/2-q_{b}^{r})/\hbareff$.  
In the semiclassical limit $x_{\text{max}}\to\infty$ and for fixed quantum 
number $m$ the integral in 
Eq.~\eqref{eq:maps:sc:tunnelrate_randformel_intnorm_a},
\begin{eqnarray}
\label{eq:maps:sc:tunnelrate_randformel_integral}
I = \Int_{0}^{x_{\text{max}}} \ue^{-2x|p(q_{b}^{r})|} [1-\cos(c_b x)]\,\ud x,
\end{eqnarray} 
becomes an $\heff$-independent constant. The tunneling rate $\gamma_{m}^{r}$ is 
proportional to the square of the modulus of the regular wave 
function at the right border $q_{b}^{r}$ of the regular island. 
With Eq.~\eqref{eq:maps:sc:wkb_wf_1d_ort} we obtain
\begin{eqnarray}
\label{eq:maps:sc:tunnelrate_rand_wkb_einges}
 \gamma_{m}^{r} & \approx & \frac{I\omega \heff}{2\pi^2\vert p(q_{b}^{r})\vert} 
    \exp\left(-\frac{2}{\hbareff} \Int_{q_{m}^{r}}^{q_{b}^{r}} \left\vert p(q')
               \right\vert \text{d}q'\right).
\end{eqnarray}
A similar equation holds for $\gamma_{m}^{l}$.

As an example for the explicit evaluation of 
Eq.~\eqref{eq:maps:sc:tunnelrate_rand_wkb_einges} 
we now consider the harmonic oscillator $\Hreg(p,q) = p^2/2+\omega^2q^2/2$,
where $\omega$ denotes the oscillation frequency and gives the ratio of the 
two half axes of the elliptic invariant tori.
Its classical turning points $q_{m}^{r,l} = \pm \sqrt{2E_m}/\omega$, 
the eigenenergies $E_m = \hbareff\omega(m+1/2)$, and the momentum 
$p(q) = \sqrt{2E_m-q^2\omega^2}$ are explicitly given.
Using these expressions in Eq.~\eqref{eq:maps:sc:tunnelrate_rand_wkb_einges} 
and $\gamma_m = 2\gamma_{m}^{r}$ we obtain
\begin{eqnarray}
\label{eq:maps:sc:wkb_formula}
   \gamma_{m} =  
              c\,\frac{\heff}{\beta_m} \;
              \text{exp}\Bigg(-\displaystyle\frac{2\A}{\heff}\left[\beta_m - 
              \alpha_m \ln \left(
              \frac{1+\beta_m}{\sqrt{\alpha_m}}\right)\right]\Bigg)
\end{eqnarray}
as the semiclassical prediction for the tunneling rate of the $m$th regular 
state, where $\alpha_m=(m+1/2)(\A/\heff)^{-1}$, $\beta_m = \sqrt{1-\alpha_m}$, 
and $\A$ is the area of the regular island. The exponent in
Eq.~\eqref{eq:maps:sc:wkb_formula} was also derived in
Ref.~\cite{Mou2009}. The prefactor 
\begin{equation}
\label{eq:maps:sc:wkb_formula_prefactor}
 c = \frac{I}{\pi^2}\sqrt{\frac{\pi\omega}{\A}} 
\end{equation}
can be estimated semiclassically by solving the integral,
Eq.~\eqref{eq:maps:sc:tunnelrate_randformel_integral}, 
for $x_{\text{max}}\to\infty$.
For a fixed classical torus of energy $E$ one obtains
\begin{eqnarray}
\label{eq:maps:sc:wkb_formula_prefactor_int}
  I & \approx & \frac{1}{2|p(q_{b}^{r})|}-\frac{2|p(q_{b}^{r})|}
                 {4|p(q_{b}^{r})|^2+c_b^2}.
\end{eqnarray}
With this prefactor the prediction 
Eq.~\eqref{eq:maps:sc:wkb_formula} gives excellent agreement 
with numerically determined data over $10$ orders of magnitude in $\gamma$, 
see Fig.~\ref{fig:maps:rates_amph_ellipt2}.
For a fixed quantum number $m$ the energy $E_m$ goes to zero in the 
semiclassical limit such that one can approximate 
$|p(q_{b}^{r})|\approx\omega q_{b}^{r}$
in Eq.~\eqref{eq:maps:sc:wkb_formula_prefactor_int} which does not depend on 
$\heff$.

Let us make the following remarks concerning 
Eq.~\eqref{eq:maps:sc:wkb_formula}:
The only information about this non-generic island with constant
rotation number is $\A/\heff$ as in Ref.~\cite{PodNar2003}. In contrast to 
Eq.~\eqref{eq:deriv:final_result} it does not require 
further quantum information such as the quantum map $U$.
While the term in square brackets semiclassically approaches one, 
it is relevant for large $\heff$.
In contrast to Eq.~\eqref{eq:maps:sc:tunneling_rate_2b}, where the chaotic
properties are contained in the difference $V(q)-\widetilde{V}(q)$, 
they now appear in the prefactor $c$ via the linear 
approximation of this difference.

In the semiclassical limit the tunneling rates predicted by 
Eq.~\eqref{eq:maps:sc:wkb_formula}
decrease exponentially. For $\heff\to 0$ the values $\alpha_m$ go to zero and
$\beta_m$ to one, such that $\gamma \sim \ue^{-2\A/\heff}$ remains which 
reproduces the qualitative prediction obtained in Ref.~\cite{HanOttAnt1984}. 
We find that the non-universal constant in the exponent is $2$ which is 
comparable to the prefactor 
$3-\ln 4\approx 1.61$ derived in Refs.~\cite{PodNar2003,She2005}. 
However, our result shows more accurate agreement 
to numerical data and does not require an additional fitting parameter, 
as will be shown in Sec.~\ref{sec:ex}.

\section{Applications}
\label{sec:ex}

We study the direct regular-to-chaotic tunneling process in the systems 
introduced in Sec.~\ref{sec:kicked} starting with the simplest example
$\mapho$ with a harmonic 
oscillator-like regular island. As further systems, we consider the map 
$\mapdef$ with a deformed regular region, the map $\mapsi$ with a regular 
stripe, and finally the standard map which shows a generic 
mixed phase space.

The map $\mapho$ has a particularly simple phase space structure
with a harmonic oscillator-like regular island with elliptic invariant tori and 
constant rotation number, see 
the insets in Fig.~\ref{fig:maps:rates_amph_ellipt}. 
Numerically, we determine tunneling rates by using absorbing boundary 
conditions at $|q| \geq 1/2$.
Analytically, for the approach derived in Sec.~\ref{sec:dirtun}, 
Eq.~\eqref{eq:deriv:final_result},
we use the Hamiltonian of a harmonic 
oscillator as the fictitious integrable system
$\Ureg$. It is squeezed and tilted according to the linearized dynamics in the
vicinity of the stable fixed point located at the center of the regular island.

Figure~\ref{fig:maps:rates_amph_ellipt} shows the prediction of 
Eq.~\eqref{eq:deriv:final_result} compared to numerical data. 
We find excellent agreement over more than $10$ orders of magnitude in $\gamma$.
In the regime of large tunneling rates $\gamma$ small deviations occur which 
can be attributed to the influence of the chaotic sea on the regular states:
These states are located on quantizing tori close to the border of 
the regular island and are affected by the regular-to-chaotic transition 
region. 
However, the deviations in this regime are smaller than a factor of two.

\begin{figure}[t]
 \begin{center}
  \includegraphics[]{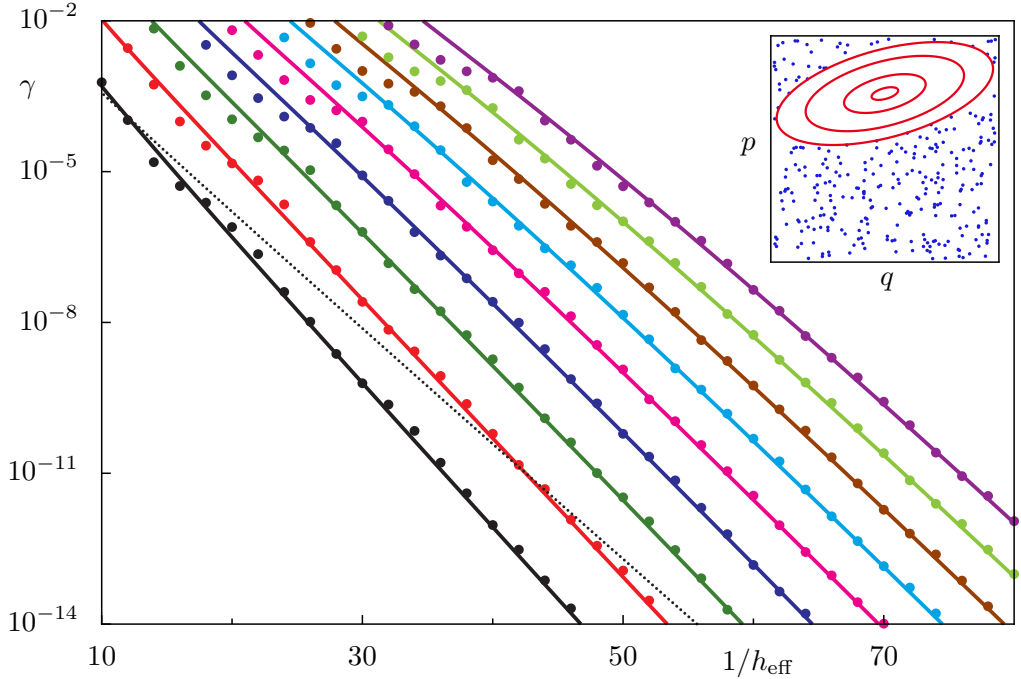}
  \caption[]
        {Numerical data (dots) for $m \leq 8$ 
         for the map $\mapho$ with a harmonic oscillator-like island.
         Comparison with Eq.~\eqref{eq:maps:sc:wkb_formula} (solid lines).
         The prediction of Refs.~\cite{PodNar2003,She2005}, 
         Eq.~\eqref{eq:maps:amphib:PodNar}, 
         for $m=0$ with a fitted prefactor is shown (dotted line).}
         \label{fig:maps:rates_amph_ellipt2}
 \end{center}
\end{figure}

In Fig.~\ref{fig:maps:rates_amph_ellipt2} we compare the results of 
the semiclassical prediction, Eq.~\eqref{eq:maps:sc:wkb_formula}, to the 
numerical data.
Due to the approximations performed in the derivation of this formula
stronger deviations are visible in the regime of large tunneling rates
while the agreement in the semiclassical regime is still excellent.

In Refs.~\cite{PodNar2003,She2005} a prediction was derived 
for the tunneling rate of the regular ground state,
\begin{eqnarray}
\label{eq:maps:amphib:PodNar}
 \gamma_0 = c \frac{\Gamma(\alpha,4\alpha)}{\Gamma(\alpha,0)},
\end{eqnarray}
where $\Gamma$ is the incomplete gamma function, $\alpha=A/\heff$,
and $c$ is a constant.
Eq.~\eqref{eq:maps:amphib:PodNar} can be approximated semiclassically 
(see Ref.~\cite{SchEltUll2005}), leading to
\begin{eqnarray}
\label{eq:maps:amphib:PodNar2}
 \gamma_0 \propto \ue^{-\frac{\A}{\heff}(3-\ln 4)}.
\end{eqnarray}
Figure~\ref{fig:maps:rates_amph_ellipt2} shows the comparison
of Eq.~\eqref{eq:maps:amphib:PodNar} (dotted line) to the numerical 
data for the map $\mapho$. Especially
in the semiclassical limit deviations are visible. The factor
$2$ which appears in the exponent of Eq.~\eqref{eq:maps:sc:wkb_formula}
is more accurate than the factor $3-\ln 4$ in 
Eq.~\eqref{eq:maps:amphib:PodNar2}.  

\begin{figure}[t]
  \begin{center}
    \includegraphics[]{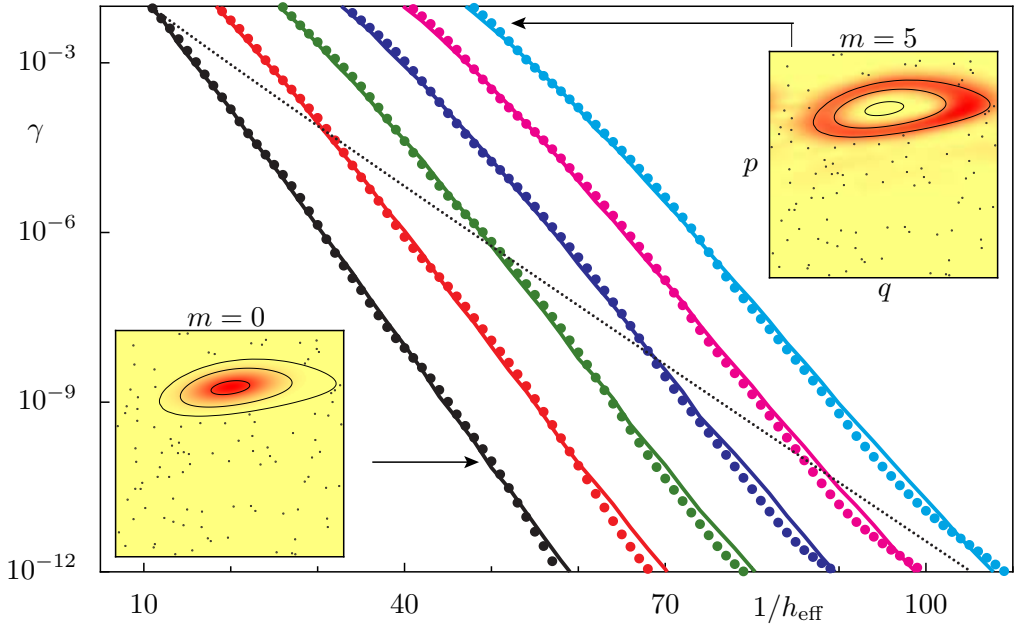}
    \caption{Dynamical tunneling rates from a regular island 
          to the chaotic sea for the map $\mapdef$: 
          Numerical results (dots) and prediction 
          following from Eq.~\eqref{eq:deriv:final_result} (lines) 
          vs $1/\hbareff$ for quantum numbers $m \leq 5$.
          The insets show Husimi representations of the regular states 
          $m=0$ and $m=5$ at $1/\heff=50$. The prediction of 
          Refs.~\cite{PodNar2003,She2005}, Eq.~\eqref{eq:maps:amphib:PodNar}, 
          for $m=0$ with a fitted prefactor is shown (dotted line).
         }
    \label{fig:maps:rates_amph_dist}
  \end{center}
\end{figure}

In typical systems the rotation number of regular tori changes from the 
center of the regular region to its border which typically has a 
non-elliptic shape. Such a situation can be achieved using the map $\mapdef$. 
Here nonlinear resonances are small such that
their influence on the tunneling process is expected only at large $1/\heff$,
see the inset in Fig.~\ref{fig:maps:rates_amph_dist} for its phase space.

We determine the fictitious integrable system $\Hreg$ by means of the 
Lie-transformation method described in Sec.~\ref{sec:dirtun:regsyst}. 
It is then quantized and its eigenfunctions are determined numerically. 
Fig.~\ref{fig:maps:rates_amph_dist} shows a comparison of 
numerically determined tunneling rates (dots) to the prediction of 
Eq.~\eqref{eq:deriv:final_result} (solid lines) yielding excellent agreement 
for tunneling rates $\gamma\gtrsim 10^{-11}$. For smaller values of $\gamma$ 
deviations occur due to resonance-assisted tunneling which is caused by a 
small $10$:$1$ resonance chain. 
Similar to the case of the harmonic oscillator-like island the fictitious 
integrable system $\Ureg$ can be approximated by a kicked
system $\Ureg\approx\UVreg\UT$ using $\widetilde{V}(q)=-rq^2/2+Rq^3/3$. 
Hence, Eqs.~\eqref{eq:maps:sc:tunneling_rate_2} and 
\eqref{eq:maps:sc:tunneling_rate_2b} can be evaluated giving similarly 
good agreement (not shown).
The prediction of Eq.~\eqref{eq:maps:amphib:PodNar} \cite{PodNar2003,She2005} 
(dotted line) shows large deviations to the numerical data.

\begin{figure}[t]
  \begin{center}
    \includegraphics[]{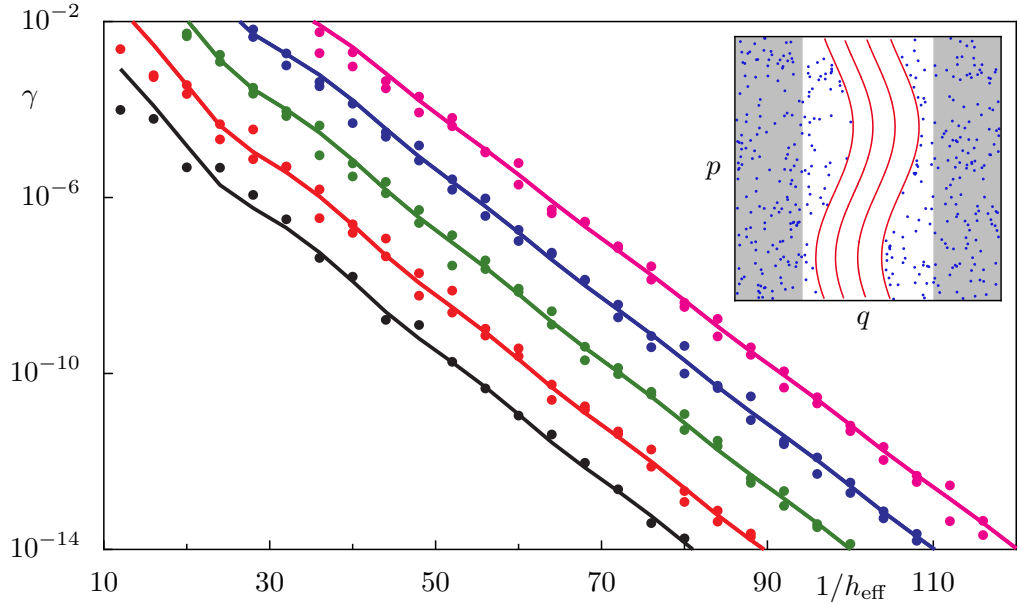}
    \caption{Dynamical tunneling rates from a regular stripe to the chaotic sea 
             for the map $\mapsi$: 
             We compare numerical results (dots) and the 
             prediction following from Eq.~\eqref{eq:deriv:final_result}
             (lines) vs $1/\heff$ for the quantum numbers $|m| \leq 4$. 
             The inset shows the phase space of the system. The numerical data
             is obtained using an absorbing region at $|q| \geq 1/4$ 
             (gray-shaded area of the inset).
          }
    \label{fig:maps:rates_shudo_map}
  \end{center}
\end{figure}

We now consider the kicked system $\mapsi$,
for which the regular region consists of a stripe in phase space, 
see the inset in Fig.~\ref{fig:maps:rates_shudo_map}. 
In Ref.~\cite{IshTanShu2007} this map is used to study the evolution of a wave 
packet initially started in the regular region by means of complex paths. 
Also for this system one can predict tunneling rates by 
means of Eq.~\eqref{eq:deriv:final_result}. 
The fictitious integrable system $\Ureg$ is determined by continuing the 
dynamics within $|q|<q_d/(8\pi)$ to the whole phase space. It is given as a 
kicked system, Eq.~\eqref{eq:maps:kicked_map}, defined by the functions
\begin{eqnarray}
 \label{eq:maps:shudo:tvreg}
 \widetilde{V}'(q) & = & -\frac{1}{2\pi}\left(\omega+\frac{d_1}{2}-
                         \frac{d_2}{2}\right),\\
 T'(p) & = & -\frac{\kappa}{8\pi}\sin(2\pi p).
\end{eqnarray}
For sufficiently large smoothing parameter $b$, see Eq.~\eqref{eq:maps:shudo:v},
the dynamics of the mixed map inside the regular 
region is equivalent to that of the regular map. Thus 
Eq.~\eqref{eq:deriv:final_result} can be applied and
we compare its results to numerically determined data.
Absorbing boundary conditions at 
$|q| \geq 1/2$ lead to strong fluctuations of the numerically determined 
tunneling rates as a function of $\heff$, presumably due to 
dynamical localization. Choosing $|q| \geq 1/4$ for the opening, which is 
closer to the regular stripe, we find smoothly decaying tunneling rates, see
Fig.~\ref{fig:maps:rates_shudo_map}. The comparison with the theoretical 
prediction shows quite good agreement.
Note, that due to the symmetry of the map there are always two regular
states with comparable tunneling rates except for the ground state $m=0$.
These two states are located symmetrically around the center of the regular 
stripe. While the theoretical prediction,
Eq.~\eqref{eq:deriv:final_result}, is identical for both of these states, the
numerical results differ slightly due to the different chaotic dynamics in the
vicinity of the left and right border of the regular region.

\begin{figure}[t]
  \begin{center}
    \includegraphics[]{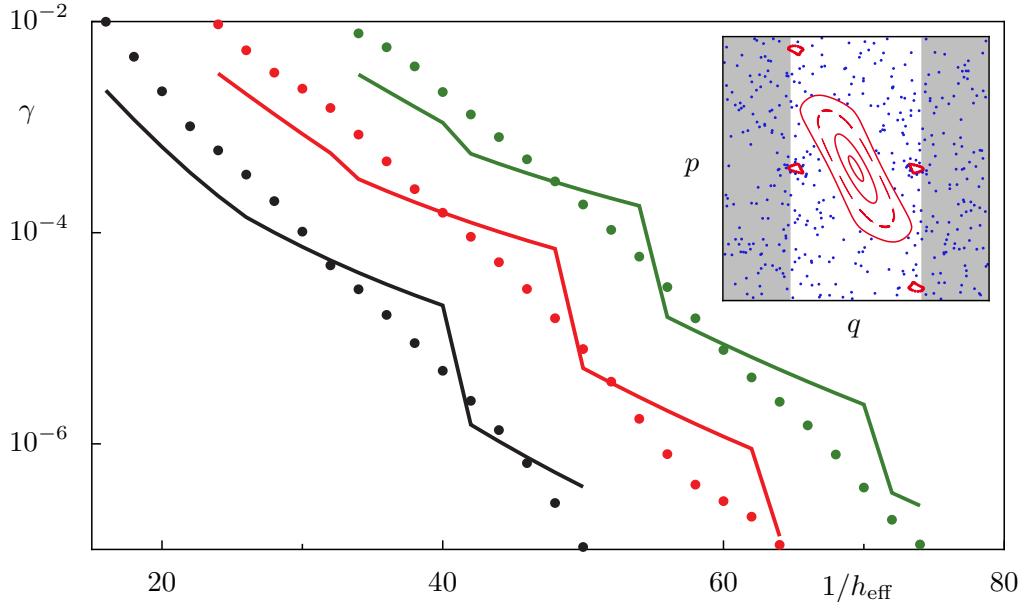}
    \caption{Tunneling rates for the standard map ($\kappa=2.9$) 
          for $m \leq 2$. 
          Prediction of Eq.~\eqref{eq:deriv:final_result} 
          (lines) and numerical results (dots), obtained using an absorbing 
          region at $|q| \geq 1/4$ (gray-shaded area of the inset). 
          }
    \label{fig:maps:rates_stmap}
  \end{center}
\end{figure}

The paradigmatic model of an area preserving map is the standard map, 
see Sec.~\ref{sec:kicked:standard}.
For $\kappa$ between $2.5$ and $3.0$ one has a large generic 
regular island with a relatively small hierarchical region surrounded by a
$4$:$1$ resonance chain, see the inset in Fig.~\ref{fig:maps:rates_stmap}. 
Absorbing boundary conditions at $|q| \geq 1/2$ lead to strong 
fluctuations of the numerically determined tunneling rates as a function of 
$\heff$, presumably caused by 
partial barriers. Choosing absorbing boundary conditions at $|q| \geq 1/4$, 
which is closer to the island, we find 
smoothly decaying tunneling rates (dots in Fig.~\ref{fig:maps:rates_stmap}). 
Evaluating Eq.~\eqref{eq:deriv:final_result} for $\kappa=2.9$ 
gives good agreement with these numerical data, 
see Fig.~\ref{fig:maps:rates_stmap} (solid lines).  
Here we determine $\Hreg$ using a method based on the frequency 
map analysis, as the Lie transformation is not able to reproduce
the dynamics within the regular island of $U$, 
see Sec.~\ref{sec:dirtun:regsyst}.  
With increasing order of the expansion series of $\Hreg$ the
tunneling rates following from Eq.~\eqref{eq:deriv:final_result} diverge. 
Hence, for the predictions in Fig.~\ref{fig:maps:rates_stmap} 
we choose terms up to second order only.
Note, that at such small order the accuracy of $\Hreg$ within the regular
region of $U$ is inferior compared to the examples discussed before. 
Hence, in Eq.~\eqref{eq:deriv:final_result} the state $U|\psireg^m\rangle$
has small contributions of all purely regular states $|\psireg^n\rangle$
in the regular island. 
These contributions are removed by the application of the projector 
$\Pch$. However, this projector depends on the number of regular states 
$\mmax$, which grows with $1/\heff$. 
If $\mmax$ increases by one, $\Preg$ projets onto a larger
region in phase space. This explains the steps of the theoretical prediction, 
Eq.~\eqref{eq:deriv:final_result}, visible in Fig.~\ref{fig:maps:rates_stmap}.

\section{Summary}
\label{sec:summary}

Dynamical tunneling plays an important role
in many areas of physics. 
Therefore a detailed understanding
and quantitative description is of great interest.
In this text we have given an overview 
on determining direct regular-to-chaotic tunneling rates
using the fictitious integrable system approach.
The direct regular-to-chaotic tunneling mechanism
is valid in the regime where Planck's constant 
is large compared to additional structures
in the regular island, such as nonlinear resonances.
To include the effect of such resonances,
resonance-assisted tunneling \cite{BroSchUll2001,EltSch2005}
has to be considered in addition with the
direct regular-to-chaotic tunneling contribution.
Recently, the two mechanisms have been studied in a combined theory \cite{LoeBaeKetSch2010}, 
giving a full quantitative description of tunneling 
from a regular island, including resonances,
into the chaotic region. 

The approach to determine direct regular-to-chaotic tunneling rates
can also be generalized to the case of billiard systems.
It can be applied if  the fictitious integrable system $\Hreg$ is known, 
as e.g.\ for the annular or the mushroom billiard. 
In addition to $\Hreg$ a description for
the chaotic states of the mixed system $H$ is needed, for which we
employ random wave models \cite{Ber1977} which account for the relevant 
boundary conditions of the billiard. For the mushroom billiard
the fictitious integrable system is easily found as the semi-circle
billiard and with this an analytical expression for the tunneling rates 
was derived in Ref.~\cite{BaeKetLoeRobVidHoeKuhSto2008}. 
This result has been compared to experimental data obtained from
microwave spectra of a mushroom billiard with adjustable stem height
and good agreement was found. 
It was also shown that tunneling rates manifest themselves in 
exponentially diverging localization lengths of 
nanowires with one-sided surface roughness
in a perpendicular magnetic field 
\cite{FeiBaeKetRotHucBur2006,FeiBaeKetBurRot2009}.

The fictitious integrable system approach has been extended to open 
optical microcavities in Ref.~\cite{BaeKetLoeWieHen2009}. 
In particular the annular microcavity was studied, which allows for 
unidirectional emission of light and shows modes of high quality factors 
simultaneously. This is desirable for most applications. 
In contrast to closed billiards the leakiness of the cavity has
to be considered, which leads to the contribution $\Qdir$ to the quality factor
caused by the direct coupling of the regular mode to the continuum. 
Additionally, the contribution caused by dynamical tunneling $\Qdyn$ 
is relevant, where $1/Q=1/\Qdir+1/\Qdyn$. $\Qdyn$ is directly 
related to the dynamical tunneling rates
given by the theory using a fictitious integrable system.
The prediction for the quality factors $Q$ has been compared to numerical data. 
Excellent agreement is found if no further phase-space structures 
exist in the chaotic sea. If additional structures appear,
the numerical data show oscillations, which cannot be explained by the
present theory. 

Future challenges include a completely semiclassical prediction of direct 
regular-to-chaotic tunneling rates in generic systems, the understanding 
of implications of additional phase space structures, and the extension 
to higher dimensional systems.

\section*{Acknowledgments}

We are grateful to S.~Creagh, S.~Fishman, M.~Hentschel, 
R.~H\"ohmann, A.~K\"ohler, U.~Kuhl, A.~Mouchet, M.~Robnik, P.~Schlagheck, 
A.~Shudo, H.-J.~St\"ockmann, S.~Tomsovic, G.~Vidmar, and J.~Wiersig
for stimulating discussions.
We further acknowledge financial support through the DFG Forschergruppe 760 
``Scattering systems with complex dynamics''.


\begin{thebibliography}{10}

\bibitem{BaeKetLoe2011}
A. B\"acker, R. Ketzmerick, and S. L\"ock:
\emph{Direct Regular-to-Chaotic Tunneling Rates Using the 
       Fictitious Integrable System Approach},
in \emph{Dynamical Tunneling - Theory and Experiment},
edited by S. Keshavamurthy and P. Schlagheck,
pp. 119 - 137, Taylor and Francis CRC (2011).

\bibitem{LanLif1979}
L.~D. Landau and E.~M. Lifschitz: {\em Lehrbuch der theoretischen Physik, Band
  3: Quantenmechanik\/}, Akademie Verlag, Berlin (1979).

\bibitem{GilPat1977}
E.~Gildener and A.~Patrascioiu: {\em Pseudoparticle contributions to the energy
  spectrum of a one-dimensional system\/}, Phys. Rev. D {\bf 16},
  423--430 (1977).

\bibitem{Per1973}
I.~C. Percival: {\em Regular and irregular spectra\/}, J. Phys. B {\bf 6},
  L229--L232 (1973).

\bibitem{Ber1977}
M.~V. Berry: {\em Regular and irregular semiclassical wavefunctions\/}, J.
  Phys. A {\bf 10}, 2083--2091 (1977).

\bibitem{Vor1979}
A.~Voros: in {\it Stochastic Behavior in Classical and Quantum Hamiltonian
  Systems}, Springer Verlag, Berlin (1979).

\bibitem{DavHel1981}
M.~J. Davis and E.~J. Heller: {\em Quantum dynamical tunneling in bound
  states\/}, J. Chem. Phys. {\bf 75}, 246--254 (1981).

\bibitem{LinBal1990}
W.~A. Lin and L.~E. Ballentine: {\em Quantum tunneling and chaos in a driven
  anharmonic oscillator\/}, Phys. Rev. Lett. {\bf 65}, 2927--2930 (1990).

\bibitem{BohTomUll1993}
O.~Bohigas, S.~Tomsovic, and D.~Ullmo: {\em Manifestations of classical phase
  space structures in quantum mechanics\/}, Phys. Rep. {\bf 223}, 43--133 (1993).

\bibitem{TomUll1994}
S.~Tomsovic and D.~Ullmo: {\em Chaos-assisted tunneling\/}, Phys. Rev. E {\bf
  50}, 145--162 (1994).

\bibitem{DemGraHeiHofRehRic2000}
C.~Dembowski, H.-D. Gr\"af, A.~Heine, R.~Hofferbert, H.~Rehfeld, and A.~Richter:
  {\em First Experimental Evidence for Chaos-Assisted Tunneling in a Microwave
  Annular Billard\/}, Phys. Rev. Lett. {\bf 84}, 867--870 (2000).

\bibitem{SteOskRai2001}
D.~A. Steck, W.~H. Oskay, and M.~G. Raizen: {\em Observation of chaos-assisted
  tunneling between islands of stability\/}, Science {\bf 293},
  274--278 (2001).

\bibitem{HenHafBroHecHelMcKMilPhiRolRubUpc2001}
W.~K. Hensinger, H.~H\"affner, A.~Browaeys, N.~R. Heckenberg, K.~Helmerson,
  C.~McKenzie, G.~J. Milburn, W.~D. Phillips, S.~L. Rolston,
  H.~Rubinsztein-Dunlop, and B.~Upcroft: {\em Dynamical tunnelling of ultracold
  atoms\/}, Nature {\bf 412}, 52--55 (2001).

\bibitem{HanOttAnt1984}
J.~D. Hanson, E.~Ott, and T.~M. Antonsen: {\em Influence of finite wavelength on
  the quantum kicked rotator in the semiclassical regime\/}, Phys. Rev. A {\bf
  29}, 819--825 (1984).

\bibitem{PodNar2003}
V.~A. Podolskiy and E.~E. Narimanov: {\em Semiclassical Description of
  Chaos-Assisted Tunneling\/}, Phys. Rev. Lett. {\bf 91}, 263601 (2003).

\bibitem{SheFisGuaReb2006}
M.~Sheinman, S.~Fishman, I.~Guarneri, and L.~Rebuzzini: {\em Decay of quantum
  accelerator modes\/}, Phys. Rev. A {\bf 73}, 052110 (2006).

\bibitem{BaeKetLoeSch2008}
A.~B\"acker, R.~Ketzmerick, S.~L\"ock, and L.~Schilling: {\em Regular-to-chaotic
  tunneling rates using a fictitious integrable system\/}, Phys. Rev. Lett.
  {\bf 100}, 104101 (2008).

\bibitem{BaeKetLoeRobVidHoeKuhSto2008}
A.~B\"acker, R.~Ketzmerick, S.~L\"ock, M.~Robnik, G.~Vidmar, R.~H\"ohmann,
  U.~Kuhl, and H.-J. St\"ockmann: {\em Dynamical tunneling in mushroom
  billiards\/}, Phys. Rev. Lett. {\bf 100}, 174103 (2008).

\bibitem{BaeKetLoeWieHen2009}
A.~B\"acker, R.~Ketzmerick, S.~L\"ock, J.~Wiersig, and M.~Hentschel: {\em
  Quality factors and dynamical tunneling in annular microcavities\/}, Phys.
  Rev. A {\bf 79}, 063804 (2009).

\bibitem{BroSchUll2001}
O.~Brodier, P.~Schlagheck, and D.~Ullmo: {\em Resonance-assisted tunneling in
  near-integrable systems\/}, Phys. Rev. Lett. {\bf 87}, 064101 (2001).

\bibitem{BroSchUll2002}
O.~Brodier, P.~Schlagheck, and D.~Ullmo: {\em Resonance-Assisted Tunneling\/},
  Ann. of Phys. {\bf 300}, 88--136 (2002).

\bibitem{EltSch2005}
C.~Eltschka and P.~Schlagheck: {\em Resonance- and Chaos-Assisted Tunneling in
  Mixed Regular-Chaotic Systems\/}, Phys. Rev. Lett. {\bf 94}, 014101 (2005).

\bibitem{SchEltUll2005}
P.~Schlagheck, C.~Eltschka, and D.~Ullmo: {\em Resonance- and Chaos-Assisted
  Tunneling\/}, in \textit{Progress in Ultrafast Intense Laser Science I},
  edited by K. Yamanouchi, S.~L. Chin, P. Agostini, and G. Ferrante, Springer,
  Berlin (2006).

\bibitem{LoeBaeKetSch2010}
S.~L\"ock, A.~B\"acker, R.~Ketzmerick, and P.~Schlagheck: {\em
  Regular-to-Chaotic Tunneling Rates: From the Quantum to the Semiclassical
  Regime\/}, Phys. Rev. Lett. {\bf 104}, 114101 (2010).

\bibitem{Ber1978}
M.~V. Berry: {\em Regular and irregular motion\/}, Am. Inst. Phys. Conf. Proc.
  {\bf 46}, 16--120 (1978).

\bibitem{ArrPla1990}
D.~K. Arrowsmith and C.~M. Place: {\em An introduction to dynamical systems\/},
  Cambridge University Press, Cambridge (1990).

\bibitem{KayMeiPer1984b}
R.~S. MacKay, J.~D. Meiss, and I.~C. Percival: {\em Transport in Hamiltonian
  systems\/}, Physica D {\bf 13}, 55--81 (1984).

\bibitem{SchOttKetDit2001}
H.~Schanz, M.-F. Otto, R.~Ketzmerick, and T.~Dittrich: {\em Classical and
  Quantum Hamiltonian Ratchets\/}, Phys. Rev. Lett. {\bf 87}, 070601 (2001).

\bibitem{BaeKetMon2005}
A.~B\"acker, R.~Ketzmerick, and A.~G. Monastra: {\em Flooding of Chaotic
  Eigenstates into Regular Phase Space Islands\/}, Phys. Rev. Lett. {\bf 94},
  054102 (2005).

\bibitem{ShuIke1995}
A.~Shudo and K.~S. Ikeda: {\em Complex Classical Trajectories and Chaotic
  Tunneling\/}, Phys. Rev. Lett. {\bf 74}, 682--685 (1995).

\bibitem{ShuIke1998}
A.~Shudo and K.~S. Ikeda: {\em Chaotic tunneling: A remarkable manifestation of
  complex classical dynamics in non-integrable quantum phenomena\/}, Physica D
  {\bf 115}, 234--292 (1998).

\bibitem{IshTanShu2007}
A.~Ishikawa, A.~Tanaka, and A.~Shudo: {\em Quantum suppression of chaotic
  tunneling\/}, J. Phys. A. {\bf 40}, F397--F405 (2007).

\bibitem{BerBalTabVor1979}
M.~V. Berry, N.~L. Balzas, M.~Tabor, and A.~Voros: {\em Quantum Maps\/}, Ann. of
  Phys. {\bf 122}, 26--63 (1979).

\bibitem{ChaShi1986}
S.-J. Chang and K.-J. Shi: {\em Evolution and exact eigenstates of a resonant
  quantum system\/}, Phys. Rev. A {\bf 34}, 7--22 (1986).

\bibitem{SchTwo2004}
H.~Schomerus and J.~Tworzyd\l{}o: {\em Quantum-to-Classical Crossover of
  Quasibound States in Open Quantum Systems\/}, Phys. Rev. Lett. {\bf 93},
  154102 (2004).

\bibitem{Wil1986}
M.~Wilkinson: {\em Tunnelling between tori in phase space\/}, Physica D {\bf
  21}, 341--354 (1986).

\bibitem{Cre1998}
S.~C. Creagh: {\em Tunneling in two dimensions\/}, in {\it Tunneling in Complex
  Systems}, World Scientific, Singapore (1998).

\bibitem{OniShuIkeTak2001}
T.~Onishi, A.~Shudo, K.~S. Ikeda, and K.~Takahashi: {\em Tunneling mechanism due
  to chaos in a complex phase space\/}, Phys. Rev. E {\bf 64}, 025201 (2001).

\bibitem{Gus1966}
F.~G. Gustavson: {\em On constructing formal integrals of a Hamiltonian system
  near an equilibrium point\/}, The Astronomical Journal {\bf 71}, 670--686 
  (1966).

\bibitem{BazGioSerTodTur1993}
A.~Bazzani, M.~Giovannozzi, G.~Servizi, E.~Todesco, and G.~Turchetti: {\em
  Resonant normal forms, interpolating Hamiltonians and stability of area
  preserving maps\/}, Physica D {\bf 64}, 66--97 (1993).

\bibitem{Sch1988}
R.~Scharf: {\em Quantum maps, adiabatic invariance and the semiclassical
  limit\/}, J. Phys. A {\bf 21}, 4133--4147 (1988).

\bibitem{LicLie1983}
A.~J. Lichtenberg and M.~A. Lieberman: {\em Regular and Stochastic Motion\/},
  Springer, New York (1983).

\bibitem{LasFroCel1992}
J.~Laskar, C.~Froeschl\'e, and A.~Celletti: {\em The means of chaos by the
  numerical analysis of the fundamental frequencies. Application to the
  standard mapping\/}, Physica D {\bf 56}, 253--269 (1992).

\bibitem{Mou2009}
J.~L. Deunff and A.~Mouchet: {\em Instantons re-examined: Dynamical tunneling 
and resonant tunneling}, Phys. Rev. E {\bf 81}, 046205 (2010).

\bibitem{She2005}
M.~Sheinman: {\em Decay of Quantum Accelerator Modes\/}, Master Thesis,
  Technion, Haifa, Israel (2005).

\bibitem{FeiBaeKetRotHucBur2006}
J.~Feist, A.~B\"acker, R.~Ketzmerick, S.~Rotter, B.~Huckestein, and J.~Burgd\"orfer:
{\em Nano-wires with surface disorder: Giant localization lengths and quantum-to-classical crossover\/},
Phys. Rev. Lett. {\bf 97}, 116804 (2006).

\bibitem{FeiBaeKetBurRot2009}
J.~Feist, A.~B\"acker, R.~Ketzmerick, J.~Burgd\"orfer, and S.~Rotter:
{\em Nanowires with surface disorder: Giant localization length and dynamical tunneling in the presence of directed chaos\/},
Phys. Rev. B {\bf 80}, 245322 (2009).

\end{thebibliography}
\end{document}